\documentclass[conference]{IEEEtran}

\usepackage{cite}	
\usepackage{graphicx} 
\usepackage[latin1]{inputenc} 
\usepackage[T1]{fontenc} 
\usepackage{amsmath,amsfonts,amsbsy,amssymb} 
\usepackage{mathabx} 
\usepackage{amssymb} 
\usepackage{amsmath}
\usepackage{amsthm}	
\usepackage{mathrsfs}
\usepackage[nolist]{acronym} 
\usepackage{tabularx} 
\usepackage{multirow}
\usepackage{wasysym}
\usepackage{float}
\usepackage{color} 

\usepackage{enumitem} 

\usepackage[colorlinks,bookmarksopen,bookmarksnumbered,linkcolor=black,citecolor=black,urlcolor=black]{hyperref}

\usepackage{subcaption}
\usepackage[ruled,norelsize]{algorithm2e}
\begin{document}

\title{Iterative Bayesian-based Localization Mechanism for Industry Verticals}
\author{Henrique Hilleshein, Carlos H. M. de Lima, Hirley Alves, and Matti Latva-aho\\
	\IEEEauthorblockA{
		Centre for Wireless Communications (CWC), University of Oulu, Oulu, Finland\\
	}
    E-mail: \{henrique.hilleshein, carlos.lima, hirley.alves, matti.latva-aho\}@oulu.fi
}

\maketitle

\begin{abstract}
We propose and evaluate an iterative localization mechanism employing Bayesian inference to estimate the position of a target using received signal strength measurements. The probability density functions of the target's coordinates are estimated through a Bayesian network. Herein, we consider an iterative procedure whereby our predictor (posterior distribution) is updated in a sequential order whenever new measurements are made available. The performance of the mechanism is assessed in terms of the respective root mean square error and kernel density estimation of the target coordinates. Our numerical results showed the proposed iterative mechanism achieves increasingly better estimation of the target node position each updating round of the Bayesian network with new input measurements.
\end{abstract}


\section{Introduction}

Over the last few years, an explosion of Machine Type Communications (MTC) has imposed stringent requirements on the operation of 5G NR and beyond systems. In such dense deployment scenarios with thousands of machines with low computational power and limited energy availability, it becomes crucial to efficiently share meager radio resources so as to enable Ultra-Reliable Low-Latency Communications (URLLC). The position of the devices connected to the wireless network is useful information to reach such requirements.  There are various technologies for positioning systems. For instance, the GPS is the most popular technology that can effectively find the position of a thing when outdoors, but it does not work properly indoor, for this reason Indoor Positioning Systems (IPS) are needed \cite{Madigan2005}.

Physical features of the wave propagation can be used to find the position of a target, such as Received Signal Strength Indicator (RSSI), Angle Of Arrival (AOA), and Time Of Arrival (TOA). Many studies about IPS have already been carried out with those physical features, such as RSSI \cite{Rusli2016, Jang2019}, AOA \cite{Navarro2017, Wielandt2017}, and TOA \cite{Naeem2018, Gustafsson2003}. Each such feature has distinct trade-offs in terms of cost, accuracy and complexity. It is well known the RSS information is a less accurate positioning metric, though it allows for using off-the-shelf devices, for example, WiFi, bluetooth, UWB \cite{Mazuelas2009}.

Thus, RSSI becomes particularly advantageous, because it can be employed without changing the current infrastructure. To find the target position, RSS-based techniques typically use fingerprinting through offline learning \cite{Jang2019, YIU2017235} or trilateration \cite{Rusli2016, Liu2018}. The figerprinting procedure not only requires measuring the RSS from several points and maintaining large measurement data sets, but also updating the fingerprint database whenever the communication ambient changes. An indoor environment usually is dynamic, so a method like trilateration where offline learning is not needed is appreciated.

Authors in \cite{Madigan2005, carlos2019} first emphasize the importance of developing IPSs that do not rely on large measurement database, and then introduce a mechanism that uses Bayesian probabilistic models to estimate the target position without any prior knowledge of the environment. Those works estimate the position of a target through Bayesian networks represented by Directed Cyclic Graphs (DAGs). The Bayesian network inference is made by Markov Chain Monte Carlo (MCMC) technique. In this work, the same method is used to find the target position, but once the position is estimated, this information is used for future estimations. The idea is to give to the system a better accuracy and a memory of the past without storing all RSSI measurements done by the access points.
In \cite{Friedman1997}, Authors describe how to update a Bayesian network model when new measurements are acquired. Herein, we extend that model by considering a sequential approach to iteratively refine the estimation of the target node position. In fact, we use the posterior distribution of the last estimation as the prior distribution each time new measurements are received. This work aims to improve non-iterative graphical models \cite{Madigan2005, carlos2019} by applying Bayesian network with the use of previous estimations of the target's position. We found out that the use of the previous estimations improve the overall performance of the system.

The reminder of this paper is organized as follows. Section \ref{sec:pgm} introduces the probabilistic graphical models, while Section \ref{sec:bayes} details the theory behind using Bayesian networks to model such localization problems. Section \ref{sec:system_model} presents the deployment scenario under investigation and mechanism implemented. The simulation and results of the mechanism are presented in Section \ref{sec:performance}. Then, we draw conclusions and final remarks in Section \ref{sc:conc}. 

\section{Bayesian Inference}
\label{sec:framework}

A Bayesian Network is a statistical model that represents the interdependence between random variables using a directed acyclic graph, and allows for drawing inference about related events conditional on our prior knowledge. Based on this model, we can make predictions of how an event behaves and we can check if our assumption reflects with the real world. Moreover, when new information is acquired about an event, we can update those prior assumptions and possibly reduce the uncertainty about the event \cite{martin2016bayesian}. Bayesian inference is a useful tool dealing with IPS
because we make inferences about the target's position based on our prior knowledge of the system and check if our prior belief reflects the reality. Next, we describe Bayesian Inference can be used to carry out indoor positioning in industry vertical deployment scenarios.

\subsection{Probabilistic Graphical Models (PGMs)}
\label{sec:pgm}
PGMs describe the underlying interdependence between the random variables in a statistical model, and thus concisely represent the corresponding joint distributions relating those variables \cite{koller2009probabilistic}. In this work, we will use DAG to represent the joint distribution utilized for the positioning estimation. Note that cyclic paths that leads a node to itself are not allowed in DAGs. In a Bayesian network, each node represents a random variable (RV) that is assumed to be conditionally independent of any other node which is not a direct descendant given its own parents \cite{Friedman1997}. It means that a RV is conditionally dependent on its own parents as shown next,

\begin{equation}
f(V) = \prod_{v \in V} f(v \vert \operatorname{pa}[v]),
\label{prodRVs}
\end{equation}

\noindent where $V$ is the set of RVs of the joint distribution and $\operatorname{pa}[v]$ is the set of the parents of the RV $v$.
%
%
As a result, the conditional distribution to any RV in the graph is given by,
\begin{align}
    f \left( v \vert V \backslash v \right) &\propto  f \left( v, V \backslash v \right) \nonumber \\
        &\propto \text{ terms in } f (V) \text{ containing } v \nonumber \\
        &= f \left( v \vert \textrm{pa} [v] \right) \negthickspace\prod_{w \in \,\text{ch} [v]} \negthickspace f \left( w \vert \textrm{pa} [w] \right),
\end{align}
\noindent where $w$ is a child of $v$ and $\operatorname{ch}[v]$ is all the children of $v$.

Herein, we employ the MCMC method to carry out Bayesian inference to estimate the position of the target node, as Bayesian's analyses are usually done through the MCMC method \cite{martin2016bayesian}. The MCMC method is a generic computational approach used to sample arbitrary distributions \cite{Kroese2011} where the sampler start with some initial values based on the prior information known about the variables, and then cycles through the graph using an algorithm to simulate each variable $v$ according to its respective conditional probability distribution \cite{Madigan2005}. In this work, the MCMC is used to find the conditional distribution of each $v$ in the graph. Succinctly, the MCMC algorithm should generate a Markov chain whose limiting distribution is equal to the desired distribution \cite{Kroese2011}. We decided to use No-U-Turn Sampler (NUTS) as the MCMC algorithm because NUTS does not have random walk behavior and it is at least as efficient as Hamiltonian Monte Carlo (HMC) \cite{martin2016bayesian}. NUTS is an extension of the algorithm Hamiltonian Monte Carlo (HMC) by avoiding random walk behavior and sensitivity to correlated parameters \cite{hoffman2014no}. The MCMC sampler uses the Bayes' theorem to find the the conditional distribution of each RV, as described next.

%

\subsection{Estimating Posterior Distributions}
\label{sec:bayes}
The Bayes' theorem uses the conditional probability to create statistical models conditional on the observations \cite{martin2016bayesian}. In fact, Bayes' theorem permits updating our prior belief about the model based on more evidence provided by new information (RSS measurements). The outcome of it is a posterior, which is a probability distribution. From \cite{martin2016bayesian}, the posterior distribution is given by, 

\begin{equation}
\label{eqbayes}
f (\mathcal{H} \vert \mathcal{D}) = \frac{f (\mathcal{D} \vert \mathcal{H}) f (\mathcal{H})}{f (\mathcal{D})},
\end{equation}
\noindent where $\mathcal{D}$ is the observed data and $\mathcal{H}$ is the assumption of the system. Moreover, $f(\mathcal{H})$ represents our prior belief about the parameters values of the model before observing any data $\mathcal{D}$, and it is called prior distribution, while $f(\mathcal{D} \vert \mathcal{H})$ yields the likelihood of verifying our prior belief given the observed data $\mathcal{D}$. $f(\mathcal{D})$ is the evidence and it is used as normalization factor.

In this work, we use Bayes' theorem to find a probabilistic distribution that portrays the data received in a way that allows to estimate the position of a target through a Bayesian network. To find the posterior distribution, first it is needed to establish suitable assumptions, by choosing the prior distribution $f(\mathcal{H})$ in ($4$), that describes our prior knowledge about the RV of interest. As aforesaid, the prior distribution initializes the MCMC sampler algorithm. The number of samples and how fast the posterior distributions converges depends on both the input data and selected prior distribution \cite{martin2016bayesian}.

This work builds upon the results in \cite{Madigan2005, carlos2019}, by using the knowledge of the past in an iterative procedure so as to find better estimations with lower uncertainty. To do that, we carry out the Bayes network inference repeated times using the posterior distribution of a previous iteration as the prior distribution for subsequent updates. It is worth mentioning that new measurement data is is fed into the model at each new iteration. Thus, at every new iteration, the proposed mechanism updates the prior distribution describing each DAG node with the corresponding posterior distribution estimate from the previous iteration. It can be seem as a system that uses a feedback loop, where the posterior information is used to define the next prior distribution.

\section{Deployment Scenario and Localization Mechanism}
\label{sec:system_model}

In this section, we describe both the test scenario and localization mechanism employed to estimate the target position. 

\subsection{Evaluation Scenario and Channel Propagation Model}
\begin{figure}[!tb] 
	\centering
	\includegraphics[width=0.866\columnwidth]{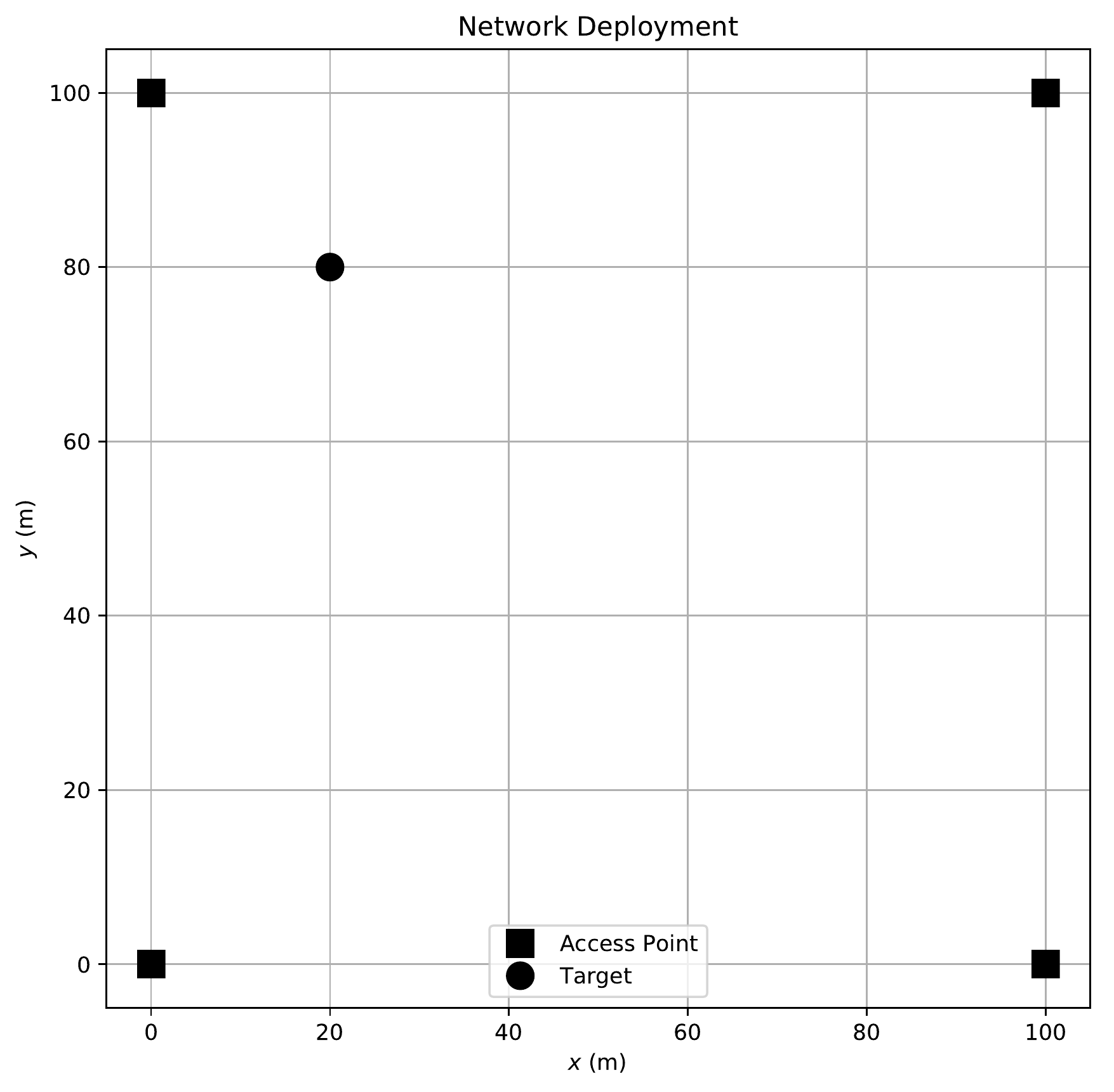}
	\caption{Illustration of the proposed scenario. Filled squares are the access points and the circle represents the target position.}
	\label{locationMap}
\end{figure}

The evaluation scenario under investigation is presented in Fig. \ref{locationMap}. It represents a squared warehouse with a side of $100$ meters. The simulation is done with four access points and each one is in the one of the corners of the warehouse, hence the location of the access points are known. Line of Sight (LOS) is assumed between the target and access points. The measurements of the RSSI made from each access points are considered uncorrelated amongst themselves.
By the combination of the known location of the access points with the assumption that the radio links are degraded by a log-distance shadowed path loss model, it is possible to estimate the position of a target with at least three access points. Note that the RSS-based localization mechanism does not require anchors to be synchronized. The radio link RSS follows a decay function given by,
\begin{equation}
\rho_i = \rho_0 - \eta \log{D_i} - \chi,
\label{eq_logdistance}
\end{equation}
\noindent where $\rho_i$ is the signal strength received at the $i$th access point, $\rho_0$ is the received power in the reference distance, in this case $1$m, $\eta$ is the path loss coefficient, $D_i$ is the euclidean distance between the target and the $i$th access point, and $\chi$ is shadowing with zero-mean normal distribution and variable standard deviation \cite{carlos2019}.

The access points send the RSSI information of the target to a server in the edge of the network. The server can estimate the position of a target after receiving a minimum amount of measurements from the access points. The estimation then can be stored to be used in the next estimation as prior knowledge of the position of the target, and the used measurements deleted. To understand how the estimation of the position was formulated, we show next the DAG representation.
\subsection{Localization Mechanism}
\label{sec:loc_mechanism}
\begin{figure}[!tb] 
	\centering
	\includegraphics[width=.7\columnwidth]{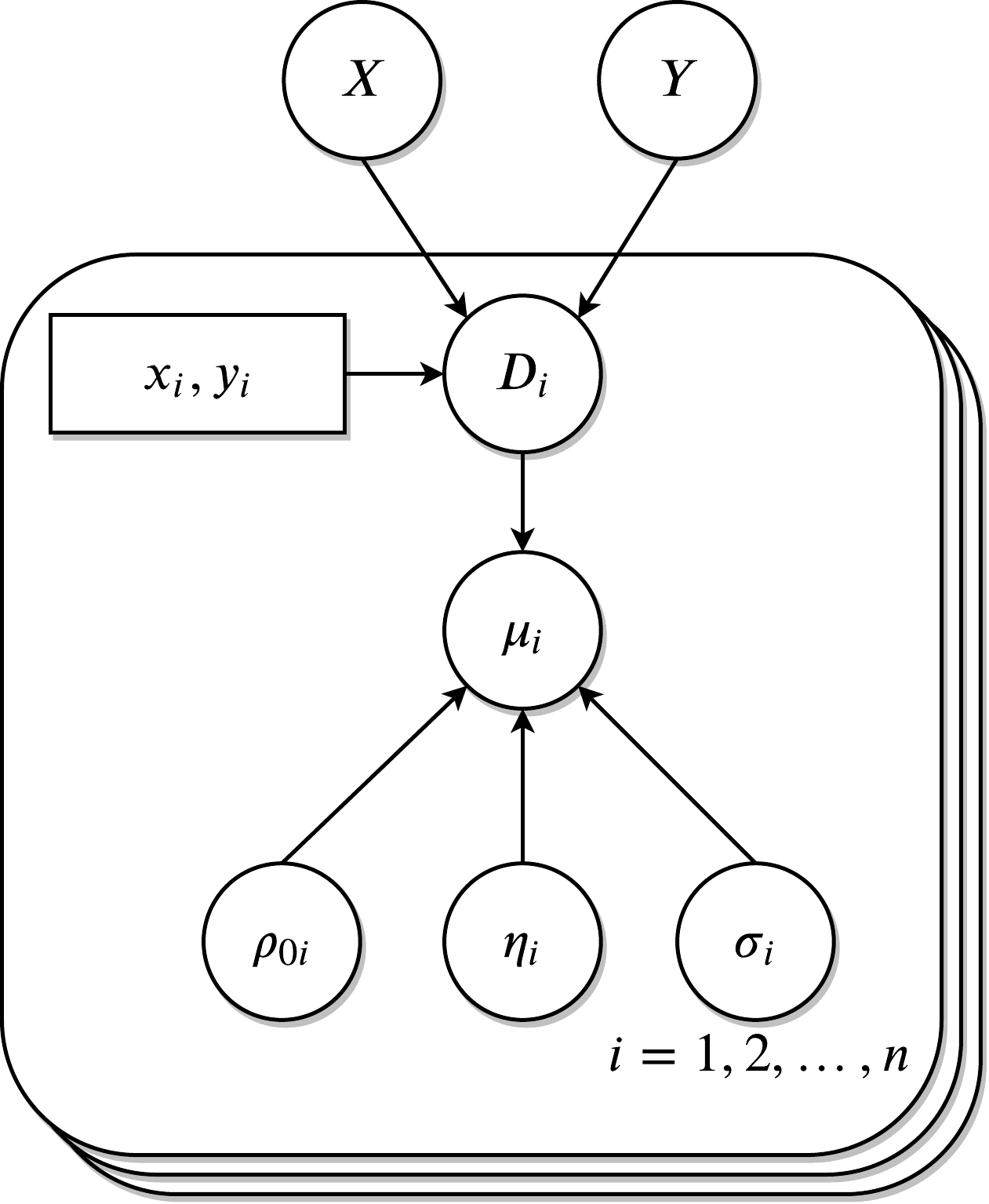}
	\caption{Bayesian probabilistic model of the RSS-based localization mechanism.}
	\label{dag_representation}
\end{figure}

The mechanism has multiple random variables whose interdependence is represented by the graphical model in Fig. \ref{dag_representation}. The symbols inside rectangles correspond to constant values, they are the coordinates of the access points. The assumptions of the random variables are based on our prior knowledge and it is represented by, 
\begin{align}
\label{eq_priorknowledge}
    &X \sim \textrm{Uniform(0, $L$)}, \nonumber \\
    &Y \sim \textrm{Uniform(0, $W$)}, \nonumber \\
    &D_i \sim \sqrt{(X - x_i)^2 + (Y - y_i)^2}, \nonumber \\
    &\mu_i \sim \rho_{0i} + \eta_i log(D_i), \\
    &\rho_{0i} \sim \textrm{Normal(0,100)}, \nonumber \\
    &\eta_i \sim \textrm{Normal(0,100)}, \nonumber \\
    &\sigma_i^2 \sim \textrm{HalfNormal(10)}, \nonumber
\end{align}
\noindent where $X$ and $Y$ are the variables that represent the target node position, $D_i$ is the distance between the target and the $i$th access point, $\rho_{0i}$ is the transmission power in an reference position (assumed to be 1m from the transmitter), $\eta_i$ is the path loss exponent and $\sigma_i$ is the standard deviation associated to the $i$th access point measurements \cite{carlos2019}.

The access points send the measurements to a server in the edge of the network, and this server runs the algorithm proposed to estimate the target's position. The first estimation (iteration) uses the first batch of measurements data acquired by the anchor nodes and applies the MCMC sampling with the mechanism described in (\ref{eq_priorknowledge}). As the position of the target is completely unknown and the target can be in any coordinate, the prior knowledge about the position is a flat distribution. The outcome of the estimation is the posteriors of the RVs. The posterior distribution is the updated belief of the system about the target's position. In this work, we use the posterior distribution when making new estimations as explained next.  

\begin{figure*} 
  \begin{subfigure}{.33\textwidth}
    \centering
    \includegraphics[width=1\columnwidth]{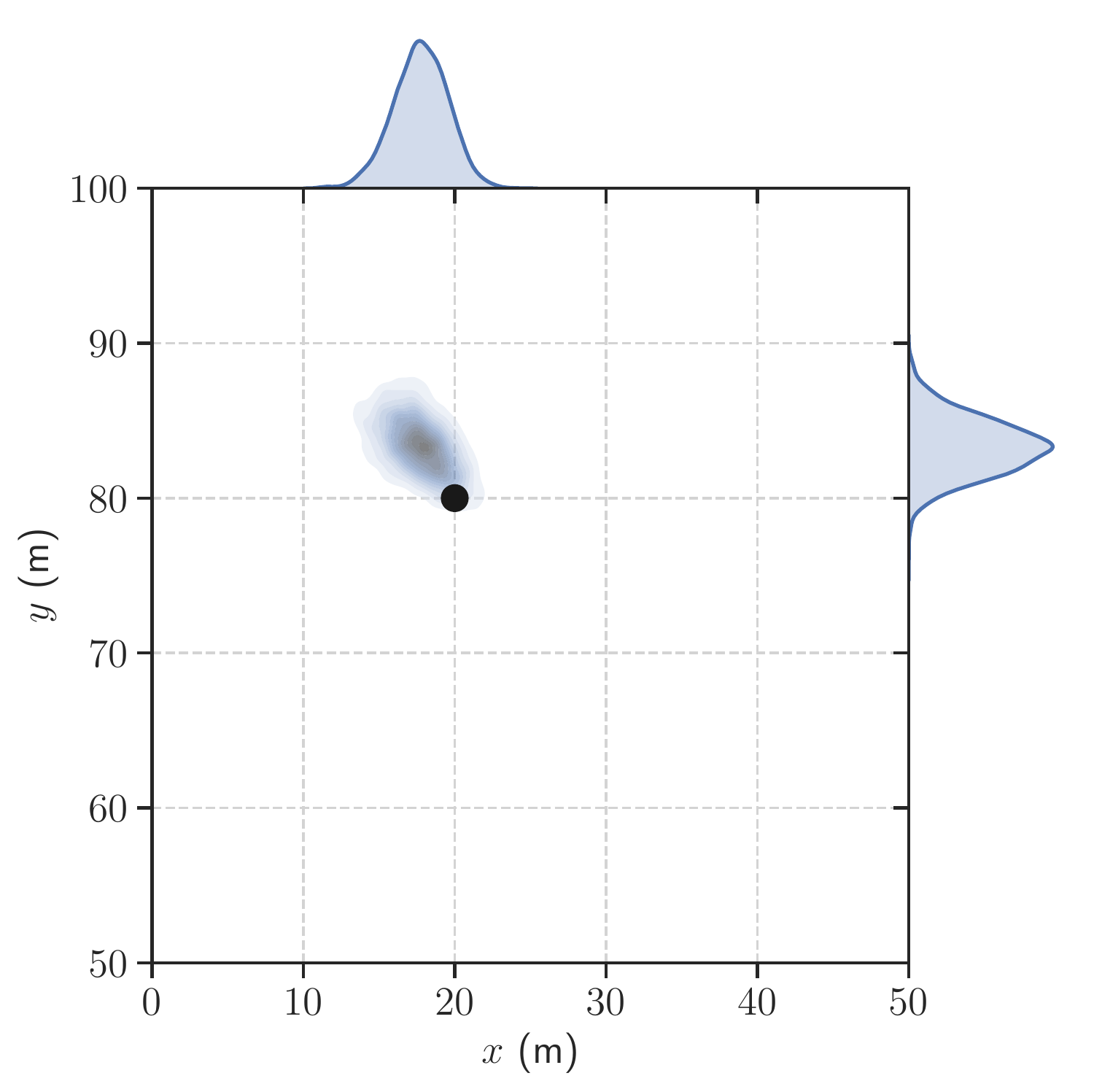}
    \caption{ }
    \label{chartPos1}
  \end{subfigure}
  \begin{subfigure}{.33\textwidth}
    \centering
    \includegraphics[width=1\columnwidth]{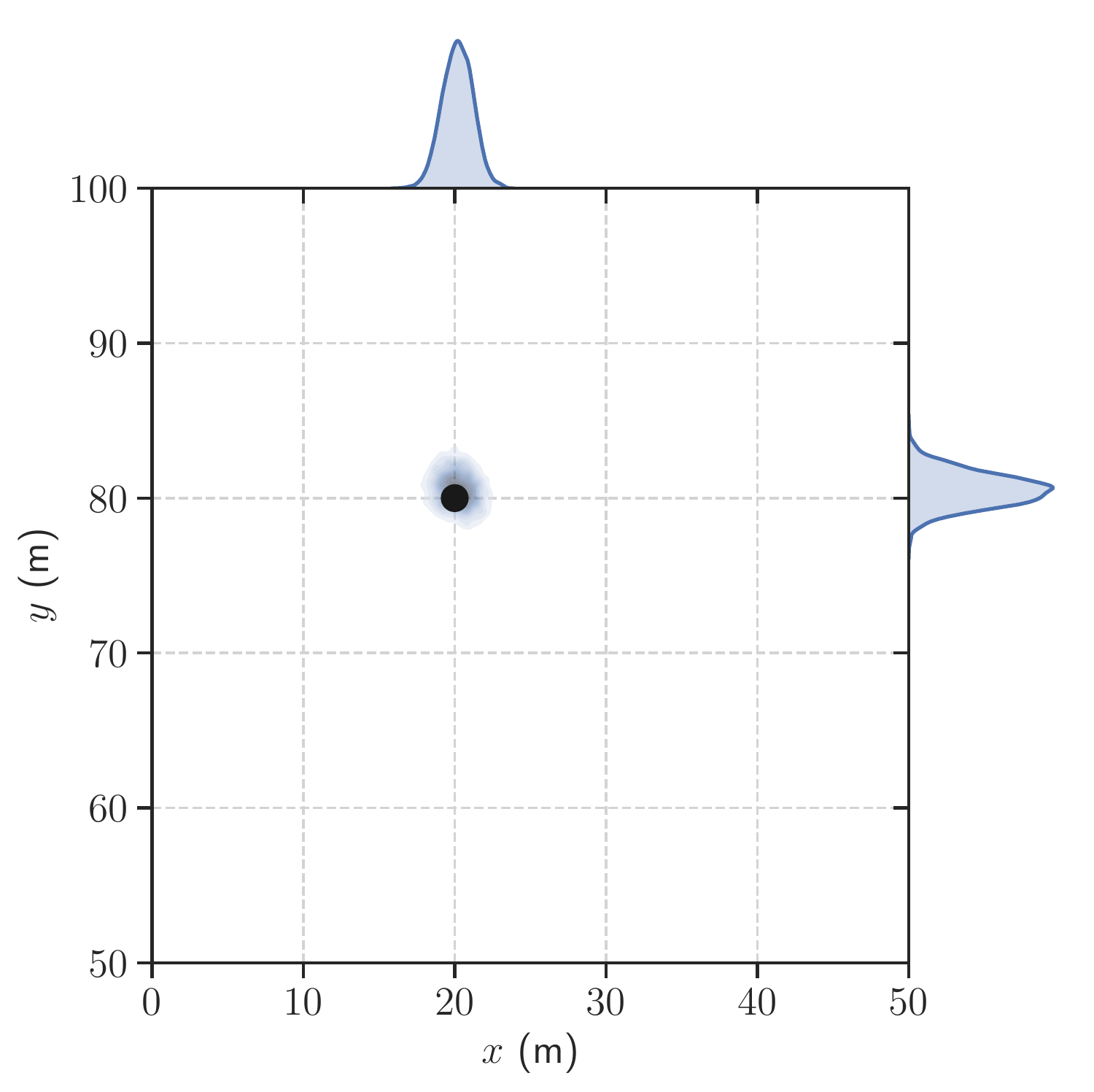}
    \caption{ }
    \label{chartPos2}
  \end{subfigure}
  \begin{subfigure}{.33\textwidth}
    \centering
    \includegraphics[width=1\columnwidth]{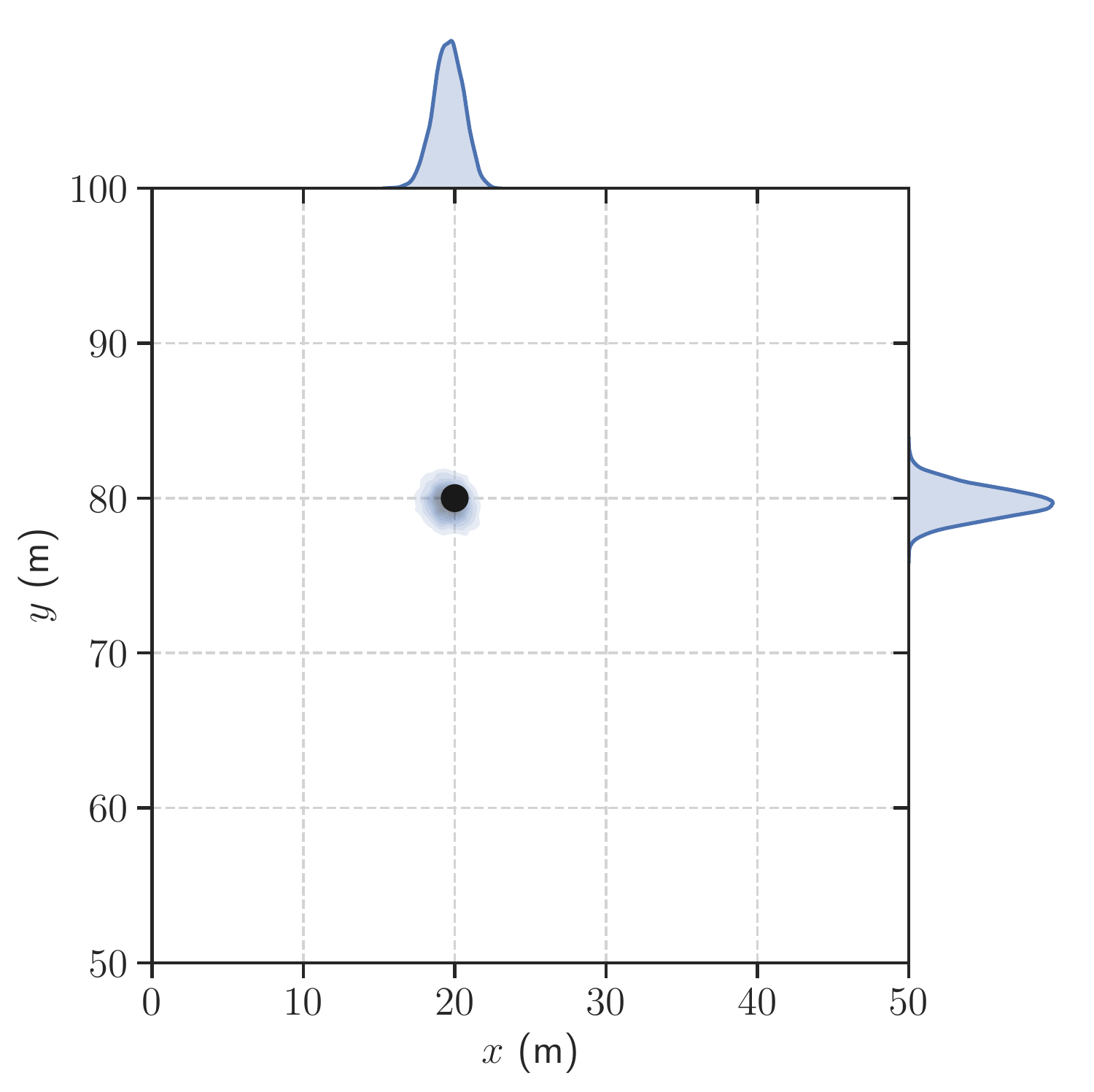}
    \caption{ }
    \label{chartPos3}
  \end{subfigure}
  \vspace{-1ex}
  \caption{Kernel density estimation for the RSS-based localization mechanism using (a) $1$, (b) $6$ and (c) $20$ iterations.}
  \label{position_estimation_graph}
\end{figure*}

\subsection{Iterative Bayesian Networks}

As mentioned before, Bayesian inference allows to make estimations based on our current knowledge about an event. When we estimate the position of a target, we have a new knowledge about the target's position, and we can use it to enhance the next estimation. We now describe the sequential update procedure used to improve the Bayesian network underlying the proposed localization mechanism. Succinctly, the posterior distribution from an arbitrary iteration is reused as the prior distribution of the following iteration. Note that the posterior distribution incorporates the evidence of previous observations (measurements statistics), therefore it reduces the need to maintain large measurements data sets.

On the other hand, a biased learning process results of reusing the posterior distribution as the prior of a subsequent iteration, which may also add error to this iterative procedure \cite{Friedman1997_2}. Indeed, the updating procedure works as a feedback loop wherein the preceding output is always used to improve our belief for the next estimation.
As previously discussed, the MCMC method provides numerical approximations to the posterior distribution, thus it is not a pure analytical approach. However, the sampling algorithm is still gradient-based, thus we resort to analytical priors (\textit{e.g.} coordinates have bivariate Uniform distribution) to initialize the underlying Bayesian network. We consider the typical approach where the measurement error follows a zero-mean normal distribution with standard deviation \(\sigma\).

The Algorithm 1 describes the mechanism algorithm, and it shows that from the second iteration and forth all RVs use its respective posterior mean and standard deviation of the previous iteration as the prior knowledge of their distribution. As the prior distribution is biased by the previous posterior, the posterior distribution obtained by the MCMC technique can converge in a wrong position, and at the same time the mechanism stop adapting to new data. It also happens when using maximum a posteriori estimation \cite{Friedman1997_2}. In this case, the standard deviation is arbitrary multiplied by two to allow the algorithm to explore the sampling space.
\begin{algorithm}[!b]
  \small
  \caption{Iterative Bayesian Network}
  \KwData{RSSI measurements}
  \KwResult{Posterior distribution of coordinates}
  initialization\;
  \While{$k < $ Max number of estimations}{
    Check measurements buffer\;
    \eIf{Data length $>=$ Minimum length}{
      Check if it is the first estimation\;
      \eIf{$k = 1$}{
        Use priors distribution of equation (5)\;
        {PosteriorDistributionsVector}[$k$] = Bayes estimation using the Data and the prior distributions\;
      }{
        Prior distributions considered to be normal\;
        The mean and standard deviation used are taken from PosteriorDistributionsVector[$k-1$]\;
        PosteriorDistributionsVector[$k$] = Bayes estimation using the Data and the prior distributions\;
      }
      $k$++\;
    }{
      Do nothing\;
    }
  }
\end{algorithm}


\section{Performance Analysis}
\label{sec:performance}
An exhaustive simulation campaign was carried out to assess the performance of the positioning mechanism. The updating prior procedure is repeated over 20 iterations. Each iteration has $250$ RSSI measurements samples of each access point. The estimation of the posterior distributions that represent the target's position ($X$, $Y$) was done using NUTS algorithm \cite{10.7717/peerj-cs.55}. The simulation follows the scenario described of a warehouse with four access points. Each access point measures the RSSI received from the target in a independent way, and the data is sent to a server in the edge of the network. When the server receives a minimum amount measurements from the access points, the estimation of the position is made using the  proposed mechanism.

\begin{figure}[!tb] 
	\centering
	\includegraphics[width=1\columnwidth]{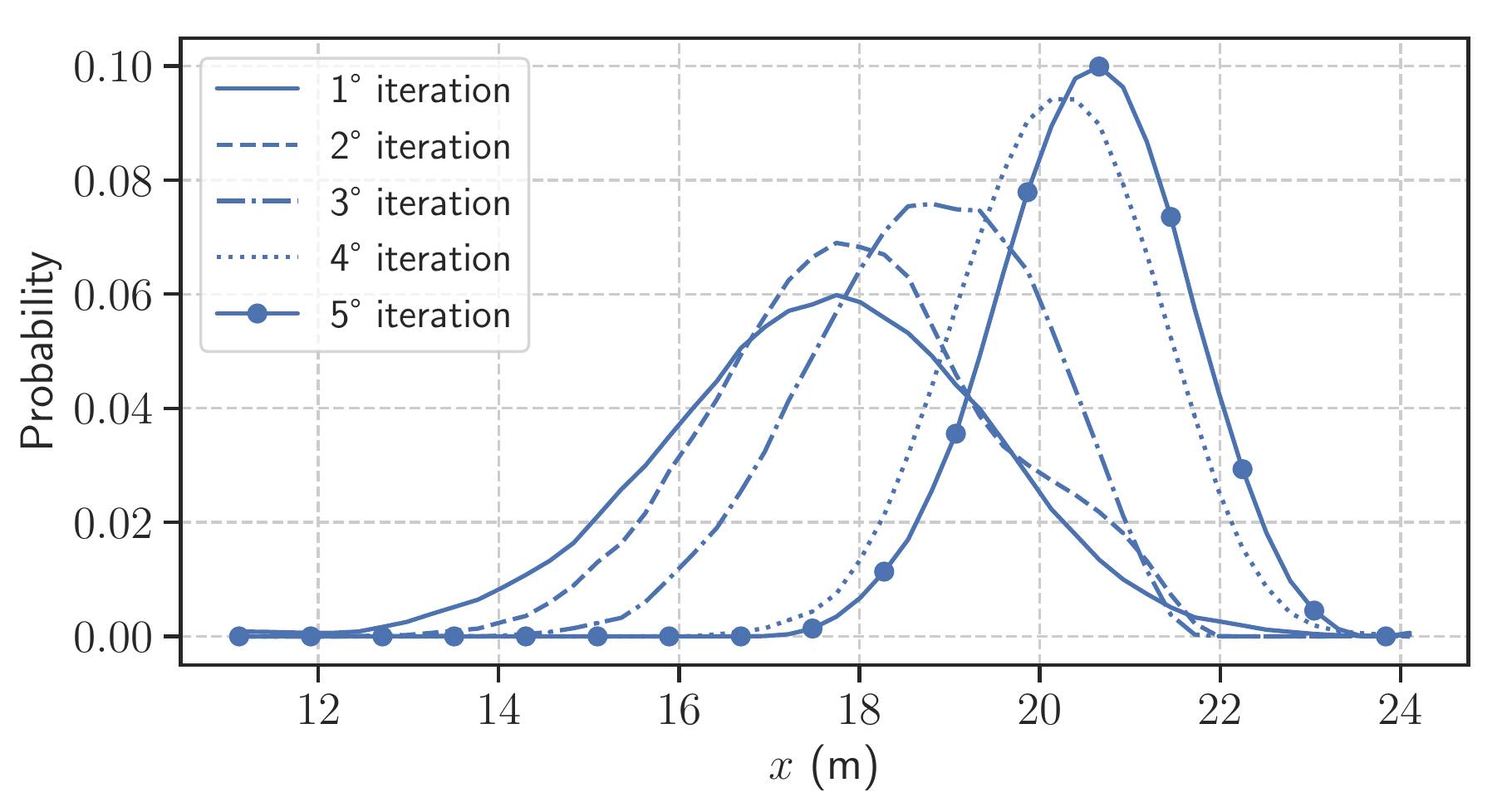}
	\caption{Posterior distribution progression of the $X$ coordinate.}
	\label{posteriorProgression}
\end{figure}

\begin{figure}[!tb] 
	\centering
	\includegraphics[width=1\columnwidth]{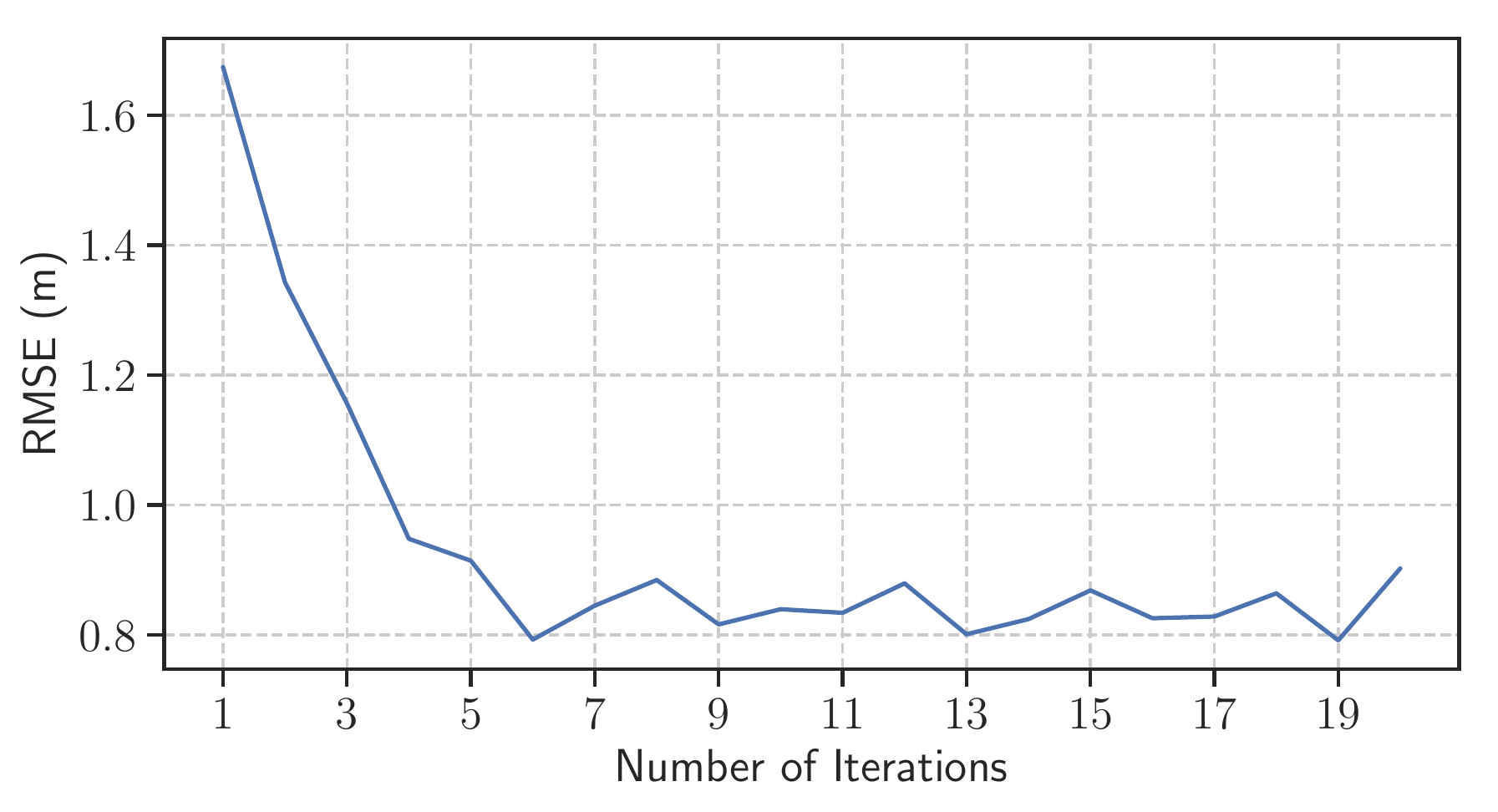}
    \caption{RMSE of the mean of $X$.}
    \label{rmsResult}
\end{figure}

Fig. \ref{position_estimation_graph} shows the outcome of the mechanism, when using (a) $1$ , (b) $6$ and (c) $20$ iterations, where $1$ iteration actually means that no update of the Bayesian network was carried out, therefore the prior distribution of the target node position is flat (no prior knowledge) as further described in (\ref{eq_priorknowledge}). Fig. \ref{posteriorProgression} illustrates the progression of the posterior distribution of the coordinate \(X\) by using the sequential update procedure. As can be seen, not only the mean value gets closer to the actual position, but also the inherent uncertainty becomes lower with more iterations. Fig. \ref{rmsResult} presents the RMSE of the mean value of the target node coordinate \(X\). This figure compares our results with the previous related works \cite{Madigan2005,carlos2019}, and shows the convergence of the system as well. In fact, the RMSE curve for one-iteration case actually corresponds to the typical non-iterative Bayesian network results. Conversely, this work considers the cases where two or more iterations are employed. After five iterations there is no significant enhancement of the estimation, so the final posterior distribution converges after five iterations. For both \(X\) and \(Y\) coordinates, the RMSE for $1$ and $6$ estimations is approximately $170$ cm and $84$ cm, respectively. The use of prior estimations provides a lower RMSE of around \(86\) cm or \(49\%\) than not using it. 

\section{Conclusions and Final Remarks} \label{sc:conc}
In this contribution, we introduce an iterative procedure based on probabilistic graphical models (Bayesian Networks) in order to estimate the target node position in the deployment scenario of interest. The proposed mechanism iterates the underlying Bayesian network by updating priors whenever new measurement data becomes available. When compared to the typical approach which does not update the prior distributions, this procedure improves the estimation. Our results show that after only five iterations, the system converges and there is no more improvement on performing further iterations.

\section*{Acknowledgments} 

The research leading to these results has received funding from the Academy of Finland through the projects 6Genesis Flagship (Grant No. \(318927\)).

\bibliographystyle{IEEEtran}
\bibliography{IEEEabrv,refs}
\end{document}